# Strain Engineering on Excitonic Properties of Monolayer GaSe

Vo Khuong Dien,[1*,] Nguyen Thi Han[1], Wei Bang-Li[1], Kuang-I Lin[2*], and Ming-Fa Lin [1,3]

[1]Department of Physics, National Cheng Kung University, 701 Tainan, Taiwan
[2]Core Facility Center, National Cheng Kung University, 701 Tainan, Taiwan
[3]Hierarchical Green-Energy Material (Hi-GEM) Research Center, National Cheng Kung University, 701 Tainan, Taiwan
*Corresponding authors: vokhuongdien@gmail.com and kilin@mail.ncku.edu.tw

# Abstract

This paper investigates strain effects on the electronic and optical properties of monolayer GaSe using first-principles calculations. The geometric deformation significantly alters energy dispersion, band gap, and the band edge states of GaSe. The band gap evolution exhibits both linearly and nonlinearly with the strains and is strongly dependent on the types of deformation and the direction of the modifications. The external mechanical strains also significantly tailor the optical properties of GaSe, the exciton binding energy is strongly reduced when the tensile strain is applied, while the opposite way is true for compressive stress. Moreover, the inhomogeneous strain also induces a non-uniform electronic screening environment and strong polarization in the absorption spectra. Our calculations demonstrate that the electronic and optical properties of GaSe monolayer can be significantly tuned by using strain engineering which appears as a promising way to design novel optoelectronic devices.

# 1 Introduction

The study of atomically thin two-dimensional (2D) has become one of the mainstream research topics of condensed matter physics due to the fascinating properties and practical applications of these materials [1-3]. The pursuit of tailoring electronic and optical properties in 2D crystals, in particular, has remained a goal in this area [4-6]. The modulations of the electronic and optical properties of 2D crystals can be established by introducing adatoms [7, 8], defects [8], mechanical deformation [9, 10], and external electric fields or magnetic fields [11-14]. Among these, strain engineering is well-known for being a useful and flexible technology to tailor material features for a variety of applications [15, 16]. For example, atomically thin carbon atoms - graphene can withstand up to 20% strain [17]. Moreover, accurate first-principles calculations predict that the broken symmetry of layered graphene under uniaxial strain induced an anisotropic Fermi velocity, significantly altering the optical properties as well as the excitonic effects [18, 19]. These findings imply that modulating the electronic structure and optical properties of graphene and other 2D systems is flexible.

A new class of 2D crystals, the atomic thins of group III-VI compounds (MX, M = Ga, In, X= S, Se, Te) have presently received a lot of attention [20-22] due to their interesting properties and their widespread use in high-harmonic generations [23, 24], optoelectronic [25-27], photovoltaics [28, 29], and photocatalytic water splitting [30, 31]. Similar to graphene, the adjacent layer of MX is weakly held together through very weak Val-der-Wall interactions, whereas the rather strong ionic interactions are dominated by the chemical bonding between post-transition metal and chalcogenide atoms in an intra-layer MX. This distinctive feature enables to cleavage of this material into a single layer via mechanical exfoliation [32, 33]. In addition, these materials can also be thinned down to the layered of the event monolayer form using other techniques. For example, a monolayer of GaSe has been successfully fabricated on the difference subtracts using chemical vapor depositions and/or molecular beam epitaxy [34-36].

According to earlier theoretical and experimental investigations, the direct-to-indirect band gap transition occurred in GaSe when the number layer is decreased below the critical value [37, 38], which is an inverse tendency of transition metal dichalcogenides [39]. In particular, up to the layer limit of 7, the lowest unoccupied state is still the parabolic sharp at the Γ point. In the contract, the significantly down-shift of the highest occupied state vicinity of the Γ point leads to a Mexican hat-sharp energy band surface around the zone center [37]. In addition to the electronic properties, layered GaSe may exhibit unique and extraordinary excitonic effects. Gallium selenide has demonstrated strong photoresponse in recent experiments [40]. The theoretical prediction of Antonius and his co-workers indicated that the absorbance spectra of GaSe exhibit feature exciton peaks with distinct polarization selectivity, such interesting phenomena come from the symmetry of the bands under in-plane mirror symmetry [41].

Up to now, there are several works have been done to investigate the effects of mechanical modulation on the stability and electronic properties of monolayer or bulk GaSe. For example, the theoretical work of Yagmurcukardes et. al has shown that such materials can be stable up to ~17% homogenous strain [42]. In addition, Z. Zhu et al predict that appropriate strain engineering can induce the Topological phase transition in layer GaSe [21]. Although these studies have been proposed inside into the effects of strain engineering; however, information about the strain effects on fundamental properties of GaSe monolayers cannot be achieved: (i) the critical mechanism, the close connection between the orbital interactions and band gap modulations has not been clarified. (ii) Changing the electronic band gap directly affects the electronic screening, many-electrons interactions in materials, and thus the excitonic effects. Studying GaSe under strain allows us to observe how the excitonic properties of deformed 2D materials to those of pristine materials. However, according to our knowledge, the delicate analysis of the inhomogeneous and homogenous strains on the optical properties and excitonic effects of group III-VI monolayers has not been established up to now.

In this work, based on the accurate first-principles density functional theory (DFT) [43] and many-body perturbation theory [44], the quasi-particle energies and the optical excitations with and without exciton effects could be obtained. By the delicate connections of the optimal geometric structure with the position-dependent chemical bonding, the band edge states in the electronic band structure, the van Hove singularities in density of states, the optical absorbance spectra with includes the excitonic effects, and the exciton wave functions, the concise picture of inhomogeneous and homogeneous strain effects could be achieved. Particularly, these strain-induced various absolute band-edge energies and exciton states may accommodate optical excitations for light emission or photo-voltaic applications.

# 2 Computational details

Vienna Ab-initio Simulation Package (VASP) [45] was utilized to perform the ground state and the excited state calculations of the GaSe monolayer.

## 2.1 Ground states calculations

The Perdew-Burke-Ernzerhof (PBE) generalized gradient approximation [46] was adopted for the exchange-correlation function. Projector-augmented wave (PAW) pseudopotentials are utilized to characterize the electronic wave functions in the core region [47]. The cutoff energy for the expansion of the plane wave basis was set to 500 eV. The Brillouin zone was integrated with a special k-point mesh of 35×35×1 in the Monkhorst-Pack sampling technique [48] for geometric optimization. The convergence condition is set to be $10^{-8}$ eV between two consecutive simulation steps, and all atoms were allowed to fully relax during geometric optimization until the Hellmann-Feynman force acting on each atom was smaller than 0.01 eV/Å.

## 2.2 Quasiparticles calculations

On top of Kohn Sham wave functions and the corresponding eigenvalues of the DFT level, the single-particle Green's function and the screened Coulomb interactions approach (G0W0) [49] were applied to get the exact electronic properties as well as wave functions for excitonic calculations. The plasmon-mode model of Hybertsen and Louie [49] was utilized to describe the screening effects. To ensure the accuracy of the calculations, we established convergence tests with various k-mesh and the Cutoff energy for the response functions. The calculated results (Figure S1) indicated that KPOINTS of $40 \times 40 \times 1$ and 130 eV for cutoff energy for response functions could give an adequate convergence for the quasi-particle band gap. These convergence parameters were used for all calculations of electronic and optical properties in this paper.

# 2.3 Optical properties and excitonic effects

Under the incidence of photons, the electrons in the occupied states will be vertically excited to the unoccupied ones in the quasiparticle energy spectra without carrying any crystal momentum transfer. According to Fermi's golden role [50], the probability of single-particle excitations can be expressed using the imaginary part of dielectric functions:

$$\epsilon_2(\omega) = \frac{8\pi^2 e^2}{\omega^2} \sum_{vck} |e.\langle vk|v|ck\rangle|^2 \delta\left(\omega - E_{\boldsymbol{ck}}^{QP} - E_{\boldsymbol{vk}}^{QP}\right),$$

where the intensity of each excitation peak and the available transition channels are directly related to the velocity matrix element, $|e.\langle vk|v|ck\rangle|^2$, and joined of the density of states $\delta\left(\omega - E_{\boldsymbol{ck}}^{QP} - E_{\boldsymbol{vk}}^{QP}\right)$, respectively.

Regarding the optical response beyond the independent particle approach, the electron-hole interactions were taken into account. The connection of the exciton energies $\Omega_S$ and corresponding electron-hole amplitude $|e.\langle 0|v|S\rangle|^2$ of the correlated electron-hole excitations S is obtained by solving the Bethe-Salpeter equation (BSE) [50]:

$$\left(E_{\boldsymbol{ck}}^{QP} - E_{\boldsymbol{vk}}^{QP}\right) A_{v c \boldsymbol{k}}^{S} + \sum_{v'c'\boldsymbol{k}'} \langle vc\boldsymbol{k}|K^{eh}|v'c'\boldsymbol{k}'\rangle = \Omega^S A_{v c \boldsymbol{k}}^{S},$$

where, $K^{eh}$is the kernel describing the correlated electron-hole pairs, $E_{\boldsymbol{ck}}^{QP}$ and $E_{\boldsymbol{vk}}^{QP}$, respectively, are the excitation energies for the conduction band states and the valence band states, and $\Omega^S$ is the energy of the excited state. In this calculation, the 6 highest occupied valence bands (VBs) and 4 lowest unoccupied conduction bands (CBs) are included as a basis for the excitonic states with a photon energy region from 0 eV to 5 eV. In addition, the Lorentz broadening parameter $\Gamma$ was set at 50 meV to replace the delta function.

# 3 Results and discussions

# 3.1 Geometric and electronic properties

We begin our investigation of the GaSe monolayer by looking at its geometric properties. The unit cell of single-layer GaSe contains four atoms (2 Ga and 2 Se). The Ga and Se atoms are connected by a very strong covalence bond in the order Se-Ga-Ga-Se, resulting in a trigonal prismatic arrangement. Similar to graphene, the top view of the GaSe monolayer also exhibits a hexagonal structure, and thus, its first Brillouin zone (FBZ) is also an equilateral hexagon. The predicted layer structure and the atom ordering of GaSe, which includes a highly anisotropic and non-uniform environment, agree well with the available experimental examinations [23, 51]. The homogenous strain (H-strain) could be achieved by uniform modulation of the lattice constant during structural relaxation. Similarly, to investigate single-layer GaSe with an asymmetric strain distribution parallel to armchair direction (A-strain) or zigzag direction (Z-strain), the deformed primitive cell along the armchair or zigzag direction with a fixed lattice constant of zigzag or armchair direction is required. Under a homogeneous (inhomogeneous) strain, single-layer GaSe retains (breaks) its hexagonal symmetry, thereby preserving (destroying) the hexagonal FBZ (Figures 1(a-c)).

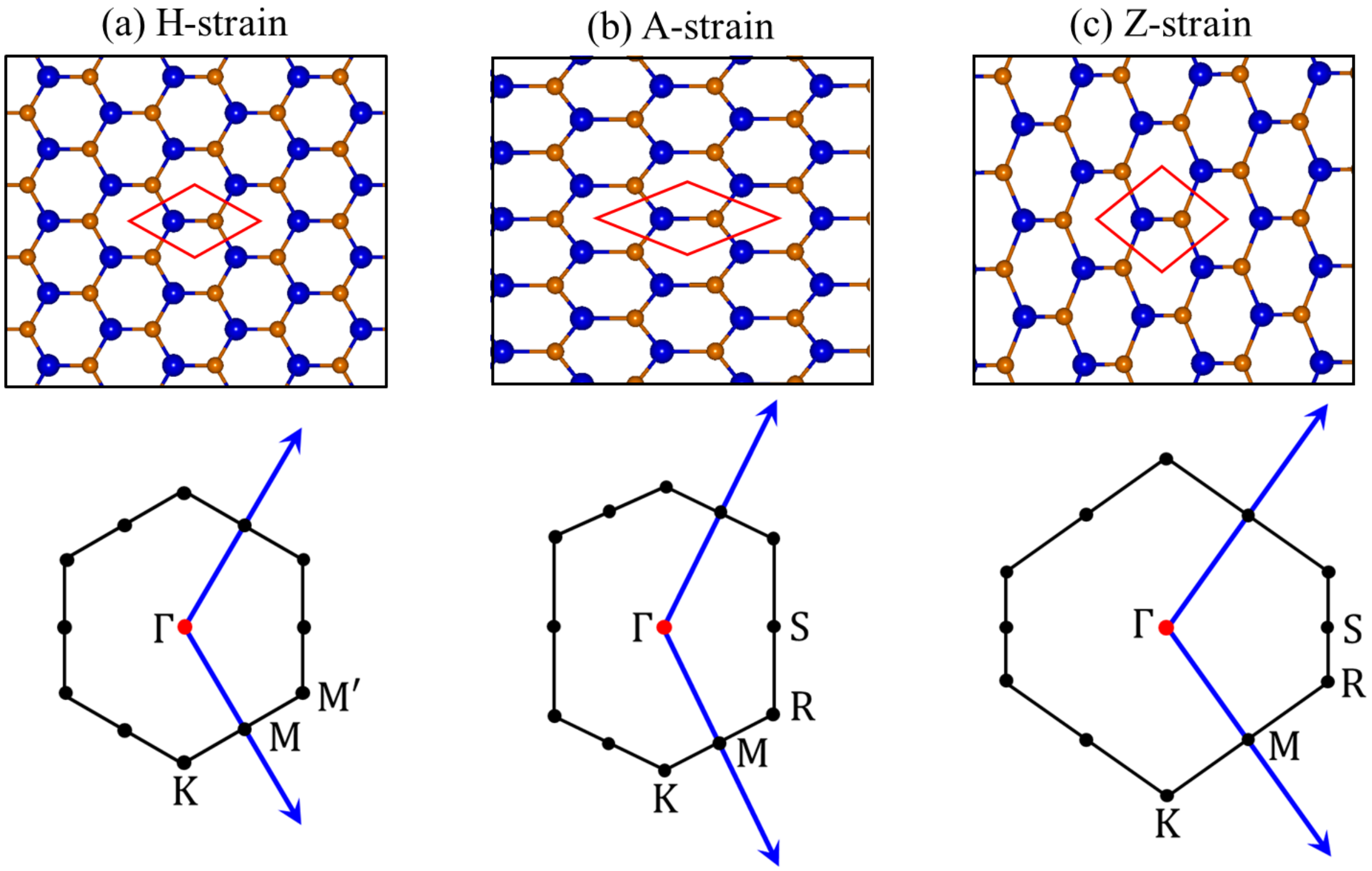

Figure 1. Structure and the FBZ of single layer GaSe. The three types of strains on monolayer GaSe and its corresponding FBZ are shown: (a) the compressive and tensile H-, (b) A-, and (c) Z-strains.

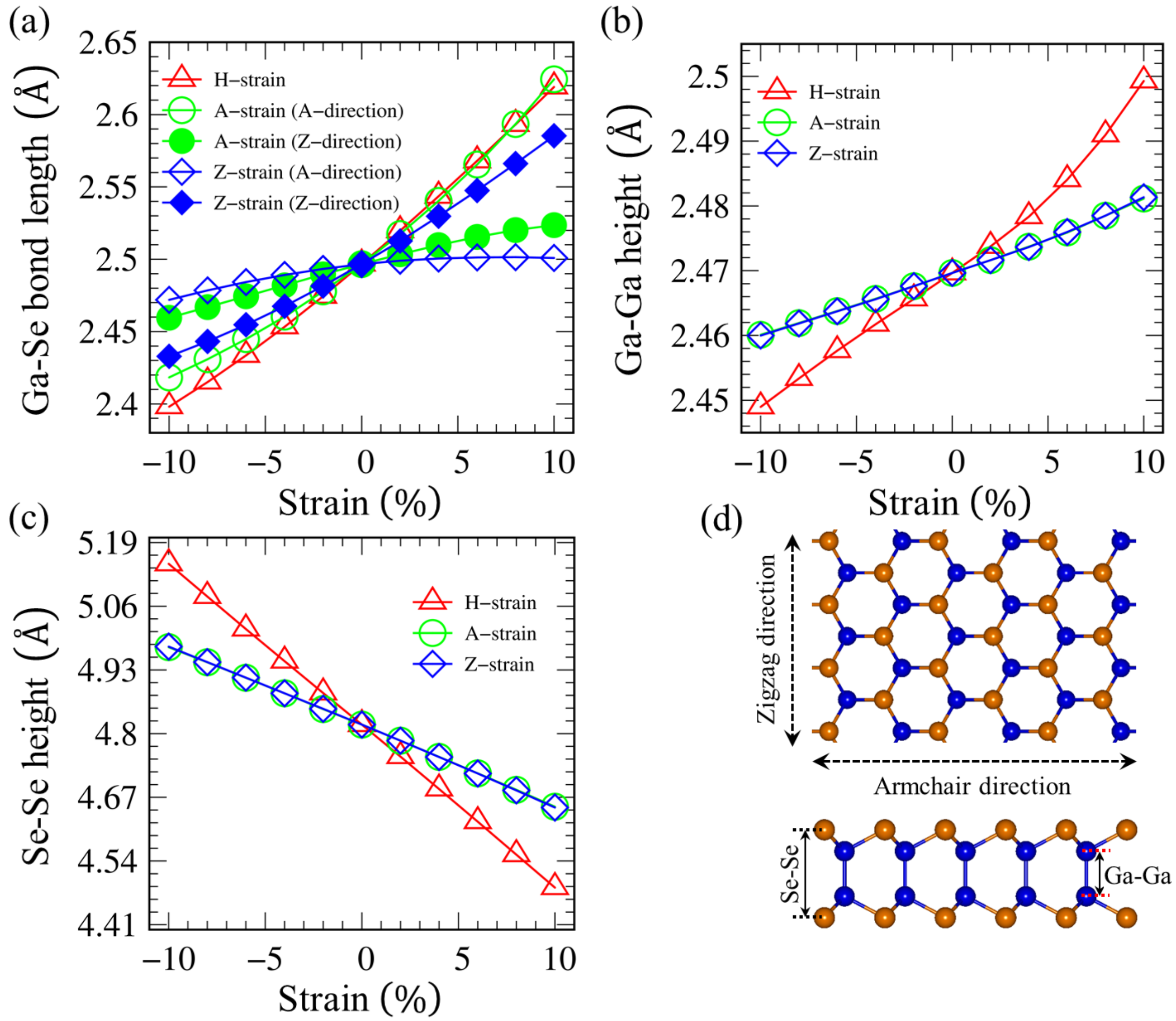

Figure 2. (a) The Ga-Se, (b) Ga-Ga bond lengths, and (c) the Se-Se height differences of GaSe monolayer under homogeneous/inhomogeneous strains. (d) Manifesting the geometric structure of GaSe monolayer.

The delicate analysis of the Se-Se height difference, Ga-Se, and Ga-Ga bond lengths under different compressive and tensile strains could provide a general picture of the strain-created deformations in monolayer GaSe. According to Figures 2(a-d), the modulations are in the following order: H-strain > A-strain ≥ Z-strain. Overall, the chemical modifications of in-plane (Ga-Se chemical bonding) and out-of-plane (Se-Se height difference) show opposite manners. The length of the Ga-Se bond increases (decreases) isotopically under tensile strain (compressive strain), the opposite trend is true for out-of-plane parameters of Se-Se height difference. The Ga-Ga chemical bonding remains almost constant under external mechanical perturbations. Although we have fixed the lattice constant along the unstrained directions, the Ga-Se bond lengths along these directions also present a slight modulation under the inhomogeneous modulations due to the Poisson effects [42, 52]. The significant chemical modifications discussed here will be directly related to the overlaps or separations of orbitals and thus, related to the modifications of electronic and optical properties.

Figure 3a depicts the electronic band structure of the GaSe monolayer. The electronic structure of GaSe has been investigated using both DFT and GW approximations, the spin-orbit coupling (SOC) is weak so it has been neglected in current calculations. The DFT level of theory indicates that the GaSe monolayer possesses an indirect gap of 1.655 eV. When the GW approximation is used, the electronic band gap increases to 3.45 eV. This value is consistent with recent GW calculations [42, 53, 54] and the STS measurement [55]. The characteristic of energy band dispersions includes the conduction band minima (CBM) $C_1$ and the parabolic relation $C_2$ at the Γ and K points, respectively. The highest occupied state presents a Mexican hat-sharp dispersion $V_1$ with its valence band maxima (VBM) located between Γ and K. The energy difference between Γ and VBM is about 0.125 eV, due to this large difference, the photoluminescence disappearance in the GaSe monolayer and other group III-VI monolayers [56]. Beyond the $V_1$ energy dispersion, the energy band structure of GaSe also be featured by two double-degeneracy $V_2$ and $V_3$, these energy bands mostly present parabolic energy dispersions. In the occupied states, the energy dispersion below the Fermi level is in good agreement with the measurement of Angle-Resolved Photoemission Spectroscopy (ARPES) as shown in Figure S2 [57].

Table 1. The optimize geometric parameters and the electronic band gap of monolayer GaSe. The previous theoretical and experimental values are also shown for comparison.

| Ga-Ga (Å) | Ga-Se (Å) | Se-Se (Å) | $E_g(eV)$ |
|---|---|---|---|
| 2.470[a] | 2.497[a] | 4.818[a] | 1.655[a]/3.446[b] |
| 2.47[c] | - | 4.82[c] | 1.83[c]/2.71[d] |
| 2.470[e] | 2.501[e] | 4.827[e] | 2.252[e] |
| 2.474[f] | 2.5[f] | - | 1.83[f] |
| - | 2.47[g] | 4.82[g] | 2.18[g]/3.68[h] |
| - | - | - | 2.0[i]/3.7[j] |
| - | - | - | 2.10[k]/3.94[l] |
| - | - | - | 3.5[m] |

a. DFT approach in this work
b. GW approach in this work
c. DFT approach in [52]
d. HSE06 approach in [52]
e. DFT approach in [37]
f. DFT approach in [58]
g. DFT approach in [42]
h. GW approach in [42]
i. DFT approach in [53]
j. GW approach in [53]
k. DFT approach in [41]
l. GW approach in [41]
m. STS measurement in [59]

The orbital-projected band structure could provide a partly information to understand the nature of electronic wave functions (Figures 3c and 3d). The delicate analysis of the orbital characteristics reveals that $V_1$, and $V_2 + V_3$ energy sub-bands, correspondingly, associate with (Ga-$4p_z$, Se-$4p_z$), and (Ga-($4p_x$, $4p_y$), Se-($4p_x$, $4p_y$)) orbitals. The features of (Ga-4s, Se-$4p_z$) and (Ga-($4p_x$, $4p_y$), Se-($4p_x$, $4p_y$)) orbitals are presented in the $C_1$ and $C_2$ states in the conduction bands, respectively. For better insight, we also calculate the band-decomposed charge density of $C_1$, $C_2$, $V_1$, and $V_2 + V_3$ energy sub-bands. As seen in Figure 3b, these sub-bands display very different spatial charge densities when compared to one another. The $V_1$ energy sub-band which dominated by the out-of-plane component of (Ga-$4p_z$, Se-$4p_z$), the $V_2 + V_3$ energy sub-bands belong to the in-plane charge density distribution of (Ga-($4p_x$, $4p_y$), Se-($4p_x$, $4p_y$)), both of them present the bonding properties. While the opposite is true for $C_1$ and $C_2$ states, these sub-bands present the anti-bonding features of (Ga-4s, Se-$4p_z$) density, and in-plane characters of (Ga-($4p_x$, $4p_y$), Se-($4p_x$, $4p_y$)) electronic density, respectively. Such identifications could be useful in explaining the band gap evolutions.

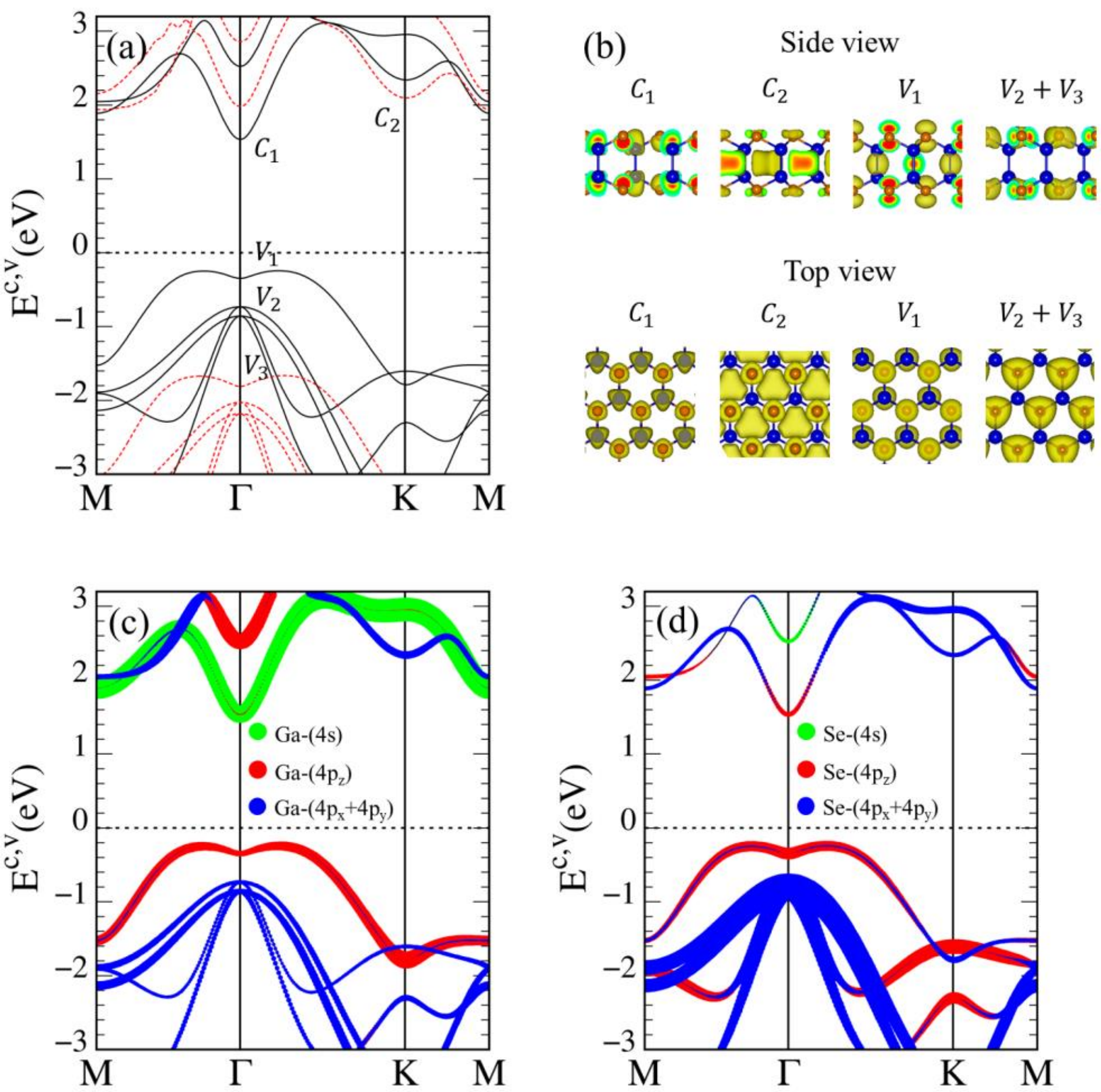

Figure 3. (a) The electronic band structure of GaSe monolayer along high symmetry points with different levels of theory. The red and black lines, respectively, indicate the GW and DFT approximations. (b) Band decomposed charge density for certain critical points in valence and the conduction bands as described in (a). The orbital projected electronic band structure for (c) Gallium atom, and (d) Selenium atom.

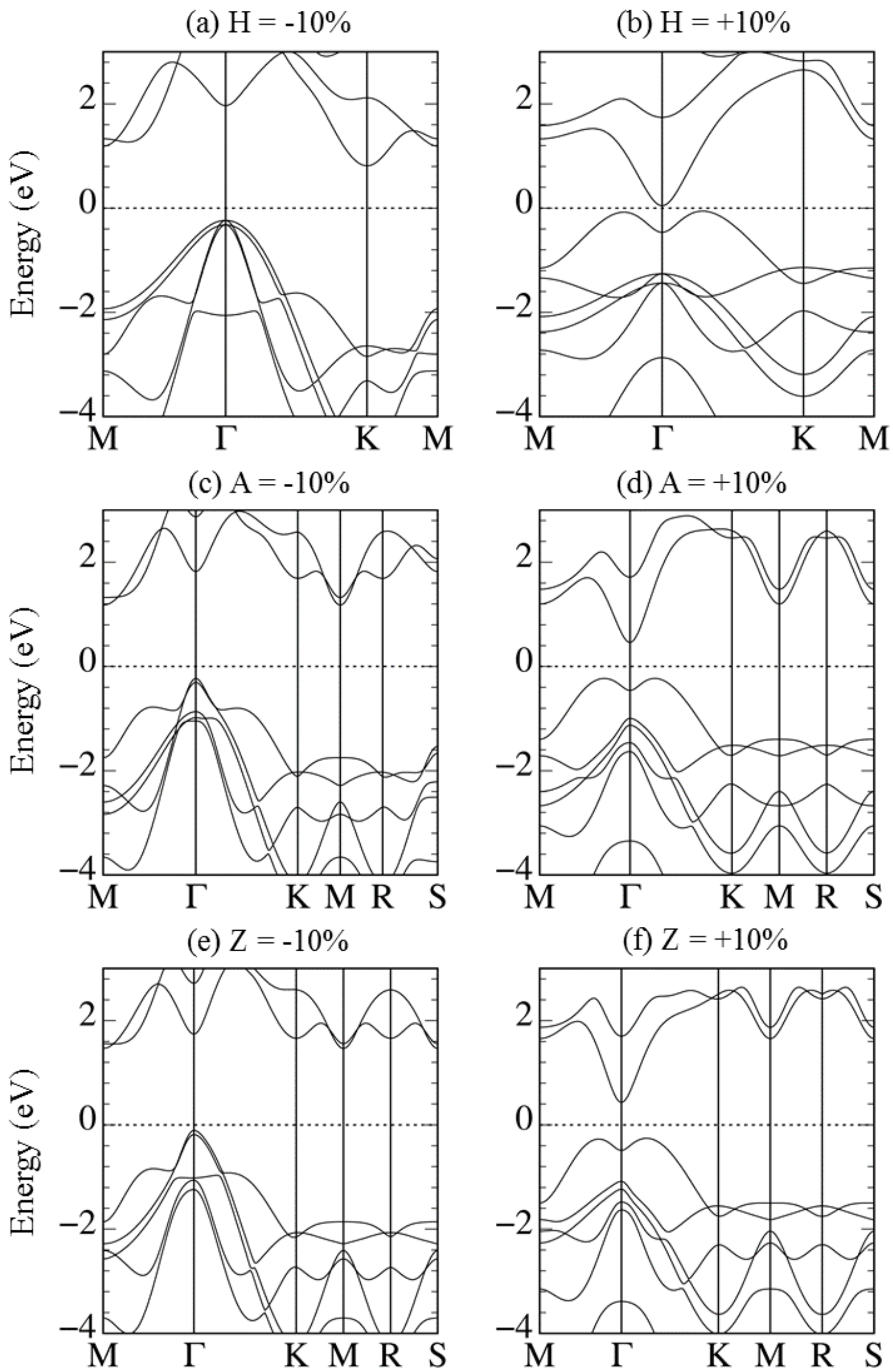

Figure 4. The DFT electronic band gap of monolayer GaSe under (a) – (b) homogeneous strains, (c) – (d) armchair strains, and (e) – (f) zigzag strains.

The modulations of the electronic band structure for GaSe upon strain are shown in Figure 4. Concerning uniform compression, the conduction energy dispersion $C_2$ and the degeneracy parabolic $V_2 + V_3$ valence sub-bands move forward to the Fermi level as a result of the Ga-($4p_x$, $4p_y$) and Se-($4p_x$, $4p_y$) orbitals overlap (resulting of Ga-Se chemical bond length decreasing). The opposite way is true for parabolic $C_1$ and Mexican hat-sharp dispersion $V_1$. Apparently, at a certain strain, the indirect-direct transition under the valence band inversion could be achieved. In the inverse process, the tensile strain induces the valence band $V_1$ and the conduction band $C_1$ to be close to each other by increasing of interactions of two-opposite Se-4$p_z$ orbitals (resulting of Se-Se height difference decreasing). Similar manners are also observed in non-uniform modulations. Very interestingly, the Mexican hat-sharp dispersion $V_1$ and the degenerate sub-bands $V_2$ and $V_3$ are either right-shift, left-shift, or vanish, and thus, the anisotropic energy band structure becomes more apparent. Furthermore, the presence of additional band edge states in specific wave-vector spaces, such as R or S points, may induce strong Van Hove singularities in the density of states, resulting in extra optical excitations.

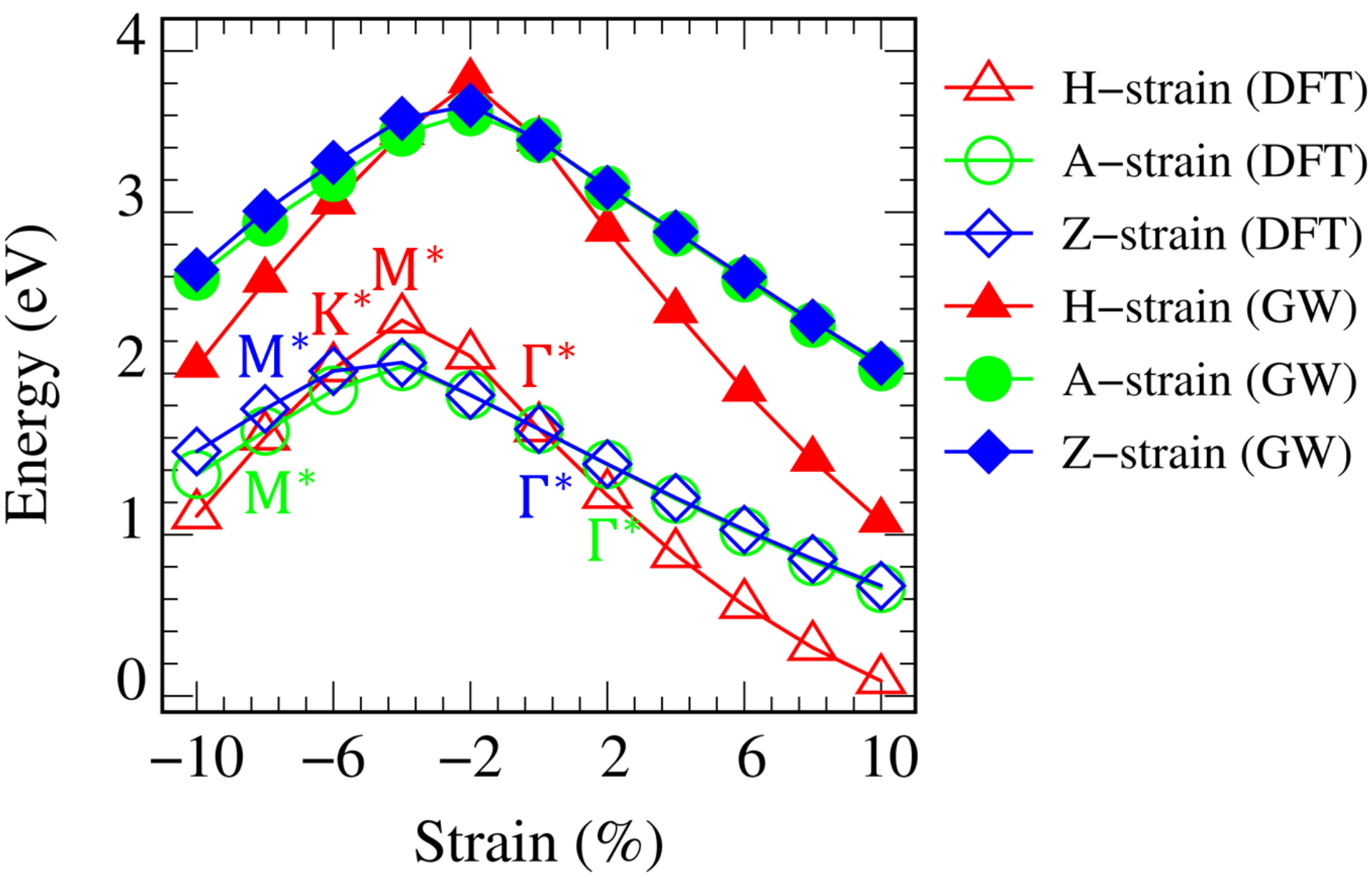

Figure 5. The band gap evolution of GaSe monolayer under homogeneous strain, armchair strain, and zigzag strain. The label $k*$ ($k = \Gamma, \mathrm{M}, \mathrm{K}$) denotes the minimum point of the conduction band, which is located at the side of the k point.

Figure 5 depict the bandgap of a GaSe monolayer as a function of external strains. It is worth noting that the bandgap adjustments for compression and elongation are significantly different. As the interactions between the atoms in the unit cell change, the energy band gap also changes significantly. Because the interaction between Ga-Se decreases with increasing H-strain, the band gap decreases linearly. The evolution of the gap for homogeneous compressive strain is more complicated; for example, it shows a slight increase to $E_g^{DFT} = 2.106\ eV/E_g^{GW} = 3.803\ eV$ and then monotonically decreases when higher strains are applied, which is due to the bottom of the conduction band transfers among several high symmetry points, such as Γ point at 0% strain, M point at -4% strain, and K point at -10% strain. Apparently, under a critical strain, the semiconductor-metallic transition could be achieved as has been illustrated in previous investigations [42]. The band gap evolution for A-strain and Z-strain also exhibits a similar tendency. However, due to the weaker atomic interactions, they present a smaller modulation. These behaviors are similar to those of monolayer $MoS_2$ [60, 61] and in good agreement with previous work [52]

# 3.2 Optical properties and excitonic effects

When the GaSe monolayer presents under the electromagnetic wave (EM), the electron will be vertically excited from the occupied states to the unoccupied ones due to the conservation of energy and momentum. The picture of the optical excitation could be well characterized by Fermi's golden rule, in which the prominence peaks in absorption spectra directly reflect the features of the electronic band structure. Beyond the single-particle picture, the mutual Coulomb interactions of the excited electron and the excited hole (excitonic effects) will significantly alter the optical absorbance spectrum, e.g., inducing the red-shift of the optical gap, and enhancement of the optical excitations [62-65]. Atomic thin 2D GaSe is expected to have strong and special excitonic effects owing to two important factors: (i) weak dielectric screening due to large quasiparticle band gap and the absence of vertical electronic screenings, (ii) quantum confinements make the excited electrons and excited holes closer. These well-defined factors will be responsible for the formation of excitonic resonances or robust exciton bound-states with remarkable binding energies. The composite effects of external strains and electron-hole interactions on the optical properties of the GaSe monolayer will be discussed in detail in the current work.

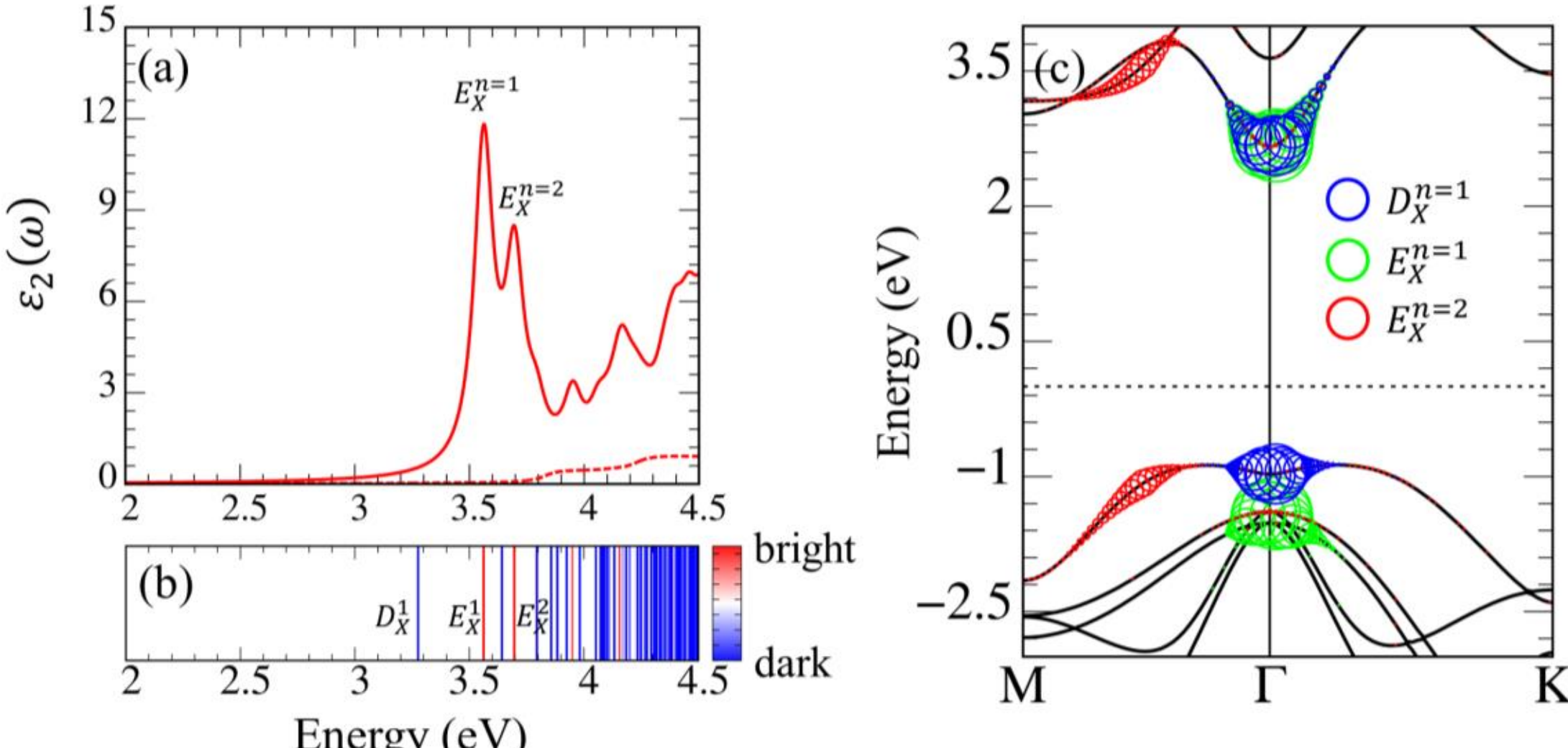

Figure 6. The optical properties of the strainless GaSe monolayer. (a) indicates the imaginary parts of dielectric functions $\varepsilon_2(\omega)$ with include (solid curves) and exclude (dashed-curves) the excitonic effects. (b) shows the energy spectrum of the bound excitons. (c) indicates the amplitude of the darkness $D_X^{n=1}$ and the first bright $E_X^{n=1}$ excitons. The radii of the circles represent the contribution of electron-hole pairs at the k-point to the $i$th exciton wave function.

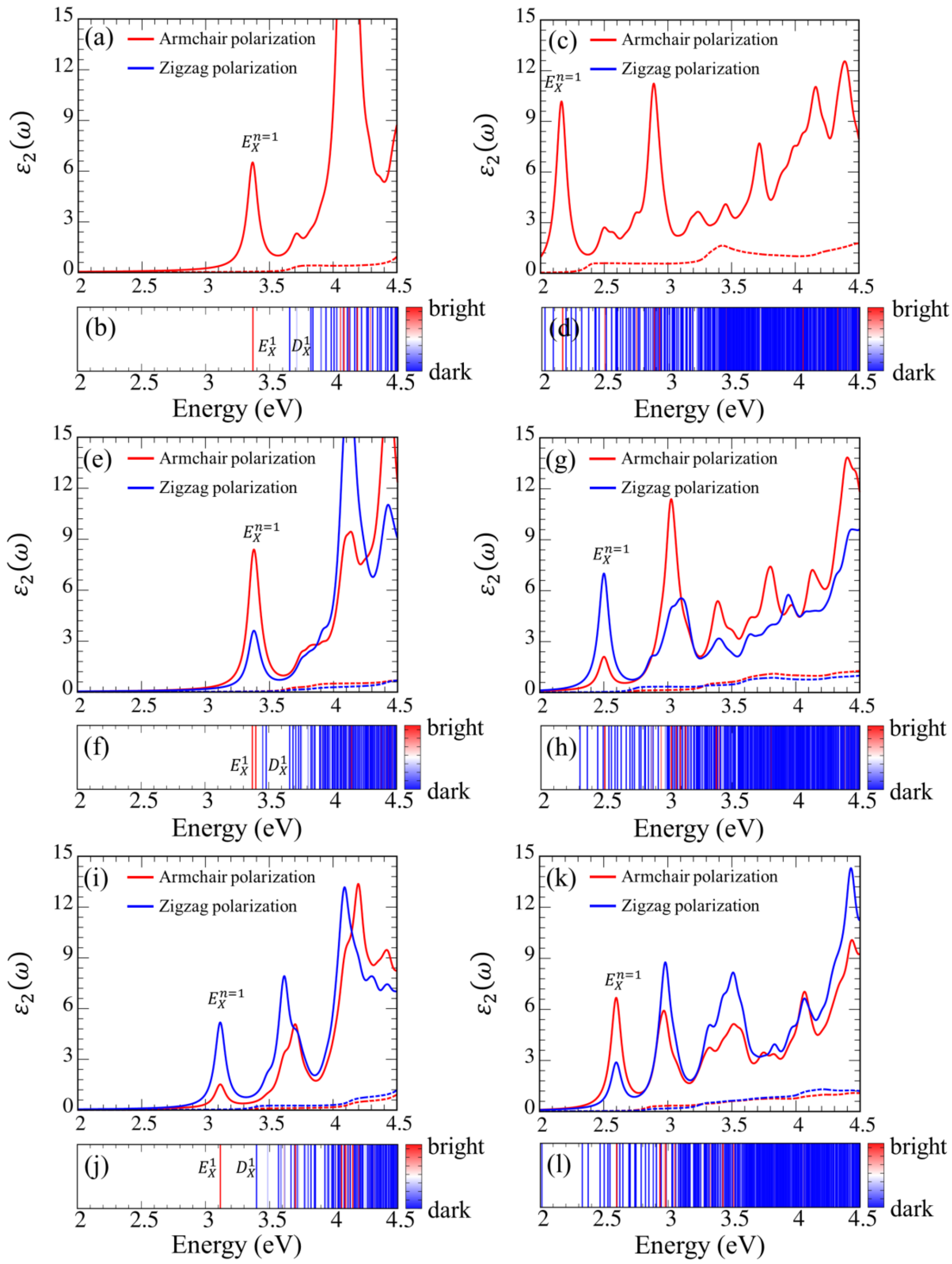

Figure 7. The optical properties of the GaSe monolayer under (a)-(b) H = −10%, (c)-(d) H = +10%, (e)-(f) A = −10%, (g)-(h) A = +10%, (i)-(j) Z = −10%, (k)-(l) Z = +10%. The top Figure indicates the imaginary parts of dielectric functions $\varepsilon_2(\omega)$ with include (solid curves) and exclude (dashed-curves) the excitonic effects. The bottom Figure shows the energy spectrum of the bound excitons.

Figure 6a shows the imaginary part of the dielectric functions ($\epsilon_2(\omega)$) of the pristine GaSe, the solid curve indicates the absorbance spectra with include the electron-hole interactions, while the dashed curve excludes these effects. Because of the symmetric of the geometric structure, the optical absorption of the GaSe monolayer is non-degenerate. The absorbance spectra in the absence of the excitonic effects in the low-frequency domain present the like-step functions, which are characteristic of two-dimensional systems with parabolic energy sub-bands. When the electron-hole interactions have been taken into account, we found two prominent bright exciton states ($E_X^{n=1}$, $E_X^{n=2}$) located below the fundamental band gap, such exciton states could be assigned to the mutual coupling of the $V_2 + V_3$ valence holes with $C_1$ conduction electrons at Γ point, and the $V_1$ valence holes with $C_2$ conduction electrons at the saddle point located between Γ and M high-symmetry direction, respectively (Figure 6c). The exciton binding energy ($E_{xb}$), which can be evaluated by the energy difference between the first bright exciton energy ($E_X^{n=1}$) and the corresponding direct transition energy is about 0.48 eV, the great red shift indicates the very strong Coulomb couplings between two opposite charges, and the composite quasiparticles in GaSe should be quite stable and comparable with other 2D systems as graphone (~0.6 eV) [66], and transition metal dichalcogenides (~1.0 eV) [67, 68]. The theoretical prediction in this work is relatively in good accordance with the previous photoluminescence measurements [27] and theoretical calculations [41].

Figure 6b shows the energy levels of the bound excitons in the GaSe monolayer including dark and bright types. Similar to other two-dimensional materials such as $MoS_2$ [69] and phosphorus [70], the exciton levels do not follow the hydrogenic Rydberg series as a result of the spatially varying screening. Even though lacking any spectral signature in the optical spectra, the dark excitons still play a significant role in understanding the optical fingerprint and the dynamics in low-dimensional materials [71]. Our calculations indicated that at least one dark exciton state is located below the first bright exciton. The minimum energy to trigger this dark state is 3.27 eV, which is 0.29 eV smaller than the required by photoexcitation. The dark exciton state with zero dipole moment for the in-plane light polarization could be understood as the forbidden transition of the opposite-parities valence band $V_1$ and the conduction band $C_1$ at the Γ center [41]. The characteristics of the bright and dark excitons are strongly modulated with the angle of the incident light as has been reported in previous work [41].

The optical excitation of GaSe under biaxial strain is shown in Figures 7(a-b) and 7(c-d), the optical gap in the absence of electron-hole interactions is reduced due to the reduction of the electronic band gap. When the excitonic effects are taken into account, the energy of prominence excitations and their intensity alter dramatically. The most prominent feature is that the exciton binding energy of GaSe under tensile strain is weak and incomparable with the strainless and the compression. The main reasons for this interesting property are due to the enhancement of the screening ability (increasing of static dielectric constant $\varepsilon_1(0)$ as shown in Table 2) or reduction of the electronic band gap and the separation of the electron and hole states. Thanks to the opposite band inversions of $V_1$ and $V_2$ sub-bands, the exciton spectra of GaSe monolayer under tensile and compressive strains show the opposite manner, there are more dark exciton states below the first bright exciton $E_X^{n=1}$ for the lattice expansions, while in the opposite case, the lowest exciton state belongs to the bright one.

The optical absorbance spectra of GaSe under armchair strain are presented in Figures 7(e-f) and 7(g-h), they also exhibit similar behaviors as in the homogeneous strains, several alien peaks appear at frequencies beyond the optical gap, the most striking difference is that the broken hexagonal symmetry results in the anisotropic optical absorbance, for the armchair compressive case, the presence of a strong exciton peak is observed for the incident light polarized along the compressive direction, while the peak with the same energy but the relatively weak intensity is observed for the incident light parallel to the zigzag direction. The opposite phenomenon is true for the expansion of the lattice constant. The polarization-dependent absorption is decided by the asymmetry of the electronic wave functions and the non-uniform of the electronic screening environments; in fact, the optical joint density of states of the first bright exciton state $E_X^{n=1}$ (the blue area in Figure 8) and the dielectric constant ($\varepsilon_1(0)$ discusses in Table 2 and Figure 9) of the direction parallel to the compression, respectively, are larger and smaller than those of the elongation, resulting in an increased probability of the corresponding transition.

The optical absorption spectra of GaSe under zigzag strain are presented in Figures 7(i-j) and 7(k-l), the imaginary part of the dielectric functions exhibit substantially anisotropic behaviors; reduced electron-hole interactions result in a significant red shift of prominent absorption peaks in the tensile strains; the optical absorbance dependents on the direction of the polarization of the incident light and no longer a constant after electron-hole interactions are included. The intensity of the exciton peak for light polarized along the armchair (zigzag) is relatively lower than that of zigzag (armchair) directions under compressive (tensile) strains, respectively. This is the inverse process of the phenomenon shown in Figures 7(e-f) and 7(g-h).

Table 2. The evolutions of the exciton binding ($E_{xb}$), and the dielectric constant $\left(\varepsilon_1(0)\right)$ under the homogeneous, armchair, and zigzag strains.

| Strain (%) | H-strain | | A-strain | | | Z-strain | | |
|---|---|---|---|---|---|---|---|---|
| | $E_{xb}$ | $\varepsilon_1(0)$ | $E_{xb}$ | $\varepsilon_1^a(0)$ | $\varepsilon_1^z(0)$ | $E_{xb}$ | $\varepsilon_1^a(0)$ | $\varepsilon_1^z(0)$ |
| -10 | 0.488 | 4.016 | 0.483 | 3.818 | 3.959 | 0.482 | 3.971 | 3.809 |
| -8 | 0.493 | 3.927 | 0.485 | 3.789 | 3.908 | 0.483 | 3.911 | 3.785 |
| -6 | 0.492 | 3.855 | 0.483 | 3.777 | 3.867 | 0.485 | 3.868 | 3.775 |
| -4 | 0.492 | 3.829 | 0.485 | 3.780 | 3.838 | 0.485 | 3.841 | 3.778 |
| -2 | 0.487 | 3.802 | 0.484 | 3.786 | 3.813 | 0.485 | 3.814 | 3.786 |
| 0 | 0.482 | 3.799 | 0.482 | 3.799 | 3.799 | 0.482 | 3.799 | 3.799 |
| +2 | 0.474 | 3.821 | 0.476 | 3.824 | 3.792 | 0.478 | 3.792 | 3.825 |
| +4 | 0.464 | 3.878 | 0.472 | 3.863 | 3.794 | 0.472 | 3.797 | 3.862 |
| +6 | 0.453 | 3.973 | 0.467 | 3.917 | 3.811 | 0.466 | 3.817 | 3.913 |
| +8 | 0.437 | 4.103 | 0.459 | 3.990 | 3.841 | 0.459 | 3.848 | 3.977 |
| +10 | 0.415 | 4.292 | 0.451 | 4.073 | 3.882 | 0.451 | 3.893 | 4.057 |

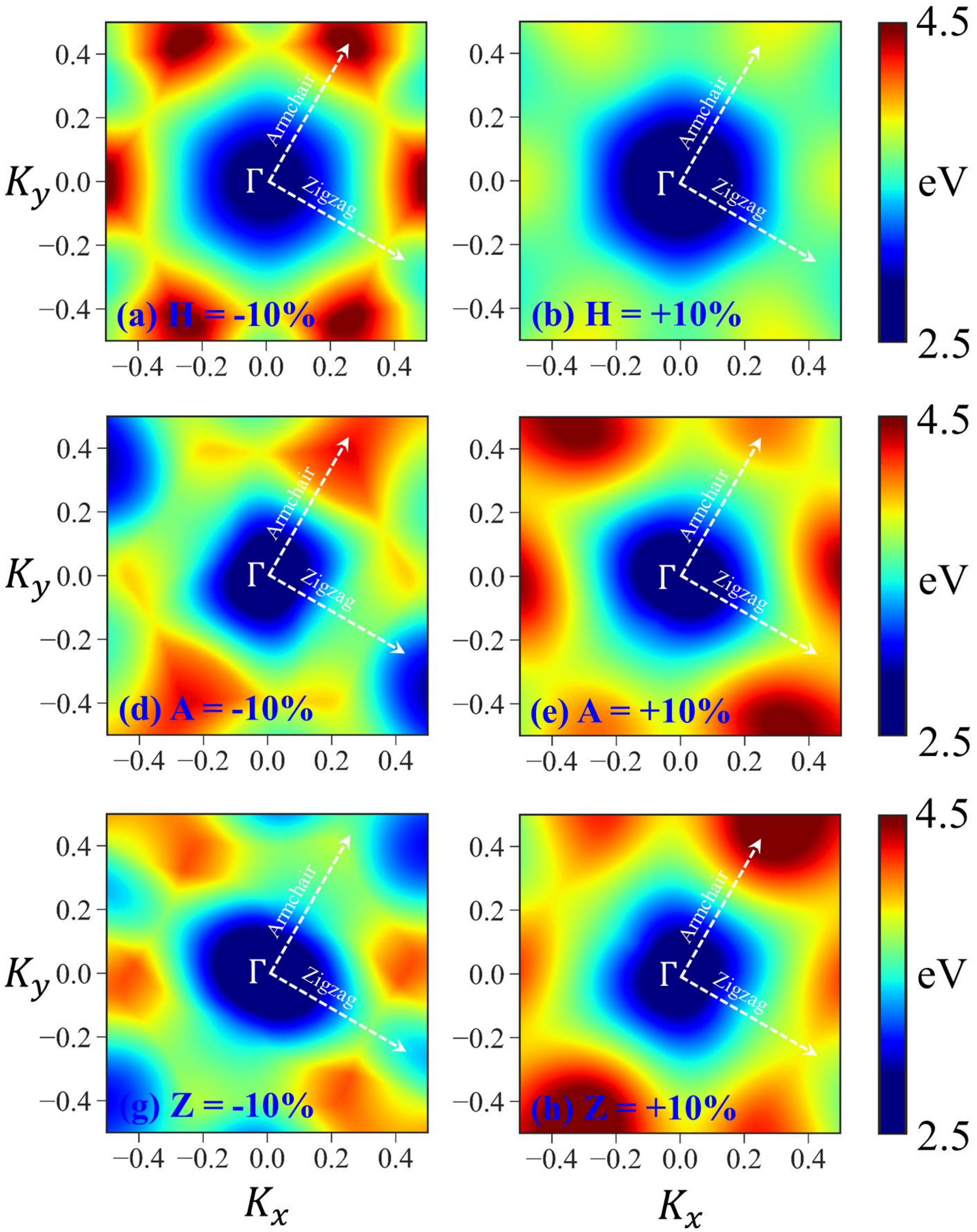

Figure 8 Direct $V_2$ valence band to $C_1$ conduction band transition energy (corresponding to $E_X^{n=1}$) in the first Brillouin zone.

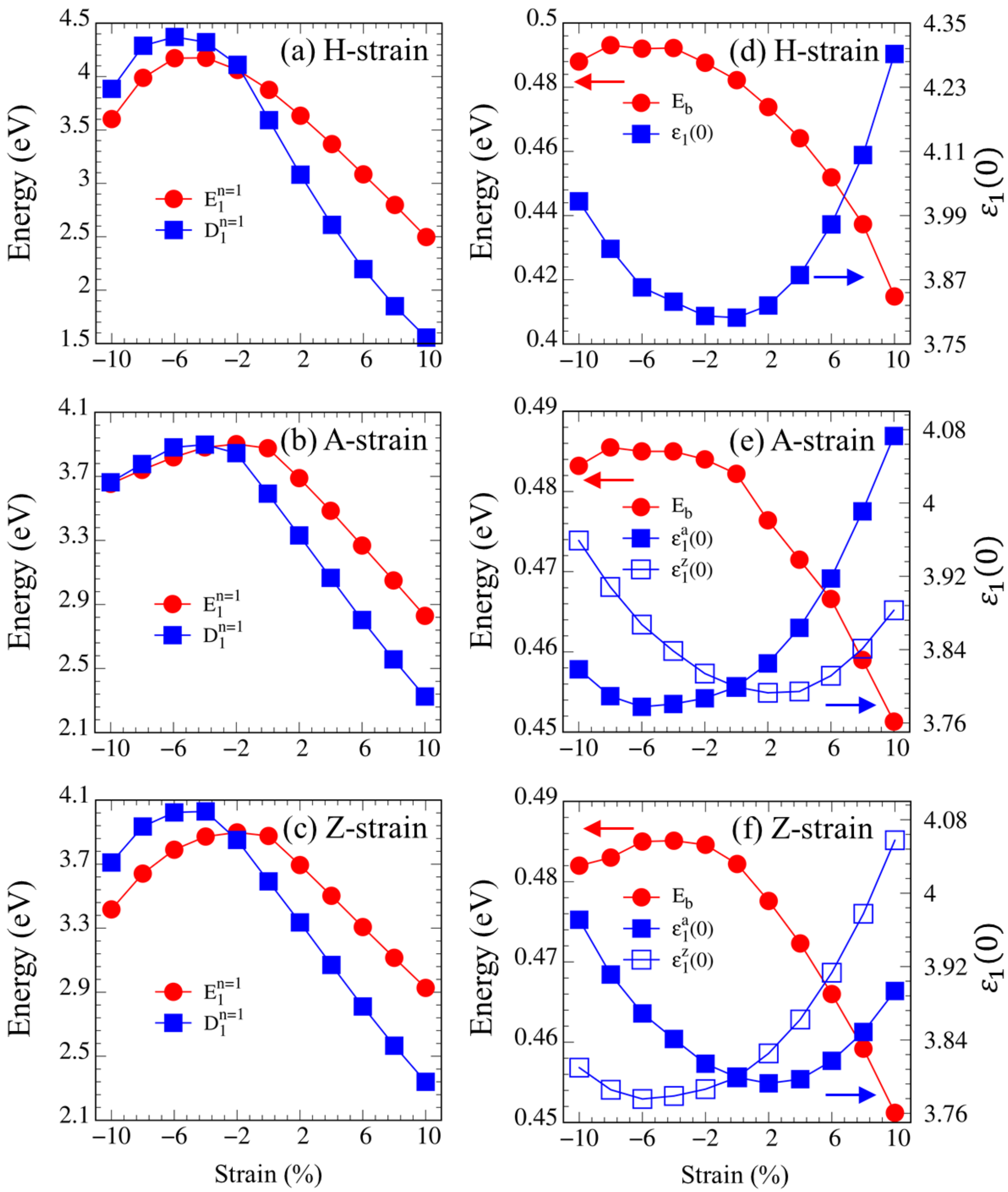

Figure 9. (a) – (c) The evolution of the first dark $D_X^{n=1}$ exciton and the first bright exciton $E_X^{n=1}$, (d) – (f) the exciton binding energy $E_{xb}$ and the static dielectric constant $\varepsilon_1(0)$ under the homogeneous, armchair, and zigzag strains.

For a better insight into the excitonic effects, the evolution of the first dark exciton $(D_X^{n=1})$, the first bright exciton $(E_X^{n=1})$, the static dielectronic screening $\varepsilon_1(0)$, and the exciton binding $E_{xb}$ have been calculated and thoroughly discussed in Figure 9.

The changing of the threshold frequency (Figure 9a) shows the different manners under compressive and tensile strains. The $E_X^{n=1}$ alters dramatically as a result of the band gap changing; it decreases linearly as the band gap decreases with increasing tensile strain. For compression, the evolution of $E_X^{n=1}$ is more complicated; for example, it increases significantly to $E_X^{n=1} = 4.18\ eV$ at H-strain = -6% and then decreases when higher strains are applied. Figure 9a also shows the evolution of the first dark exciton state $D_X^{n=1}$, there is a crossover for the dark and bright exciton types under the critical onset point H = -2%. The separation of $D_X^{n=1}$ and $E_X^{n=1}$ becomes more apparent as the strain increases towards positive values, and the dark one becomes the lowest exciton state, which agrees well with the observation that the Mexican hat-sharp dispersion becomes the top valence sub-bands and the significant energy difference between Γ and VBM. The evolution of $D_X^{n=1}$ and $E_X^{n=1}$ under A-strain (Figure 9b) and Z-strain (Figure 9c) also followed a similar trend. However, due to differences in the levels and kinds of band gap modification, they exhibit slight differences in particular characteristics, such as the critical strain and the position of $E_X^{n=1}$and $D_X^{n=1}$.

Figures 9d, 9e and 9c sheds light on the strain-dependent of the dielectric screening $\varepsilon_1(0)$, and the exciton binding energy $E_{xb}$ of the GaSe monolayer. Similar to black phosphorus [72], the exciton binding energy $E_{xb}$ of the GaSe monolayer sensitive changes under mechanical modifications. The exciton binding of the GaSe monolayer decreases linearly with increasing dielectric screening $\varepsilon_1(0)$ or decreasing the electronic band gap and well-separation of the electron and hole states, for example, $E_{xb}$ = 0.474 eV for H = +2 %, $E_{xb}$ = 0.453 eV for H = +6 %, and $E_{xb}$ = 0.38 eV for H = +10%. In contrast, $E_{xb}$ exhibits a slight increase in the external compressive strain even as the electronic screening is gradually increased, e.g., $E_{xc}$ = 0.48 eV and $\varepsilon_1(0) = 3.799$ at H = 0%, and $E_{xc}$ = 0.492 eV and $\varepsilon_1(0) = 3.855$ for H = -6%. Such a marked difference supports the notion that the physical distance controlled by mechanical strain plays an important role in determining the exciton binding energy. These findings can be interpreted qualitatively as follows: The Coulomb interaction between the negatively charged electrons and the positively charged holes, as depicted in the simple hydrogenic model $H_{hydrog} = -\frac{\nabla^2}{2m^*} + \frac{e^2}{\varepsilon.r}$, is highly influenced by the environment's electronic screening $\varepsilon$. The $E_{xb}$, on the other hand, is determined by the overall effects of the electronic screening as well as the physical distance between the excited hole and the excited electron r (Schematics in Figure 10). Figure 9d indicates that the effects of the distance between two opposite-charge particles have a greater influence on their attractive couplings than that of $\varepsilon_1(0)$, and as a consequence, the $E_{xb}$ gradually increases with compressive strain. Not surprisingly, the impact of electron-hole physical distance no longer surpasses that of the environmental screening, and the $E_{xb}$ starts to diminish at the critical strain H = -8%.

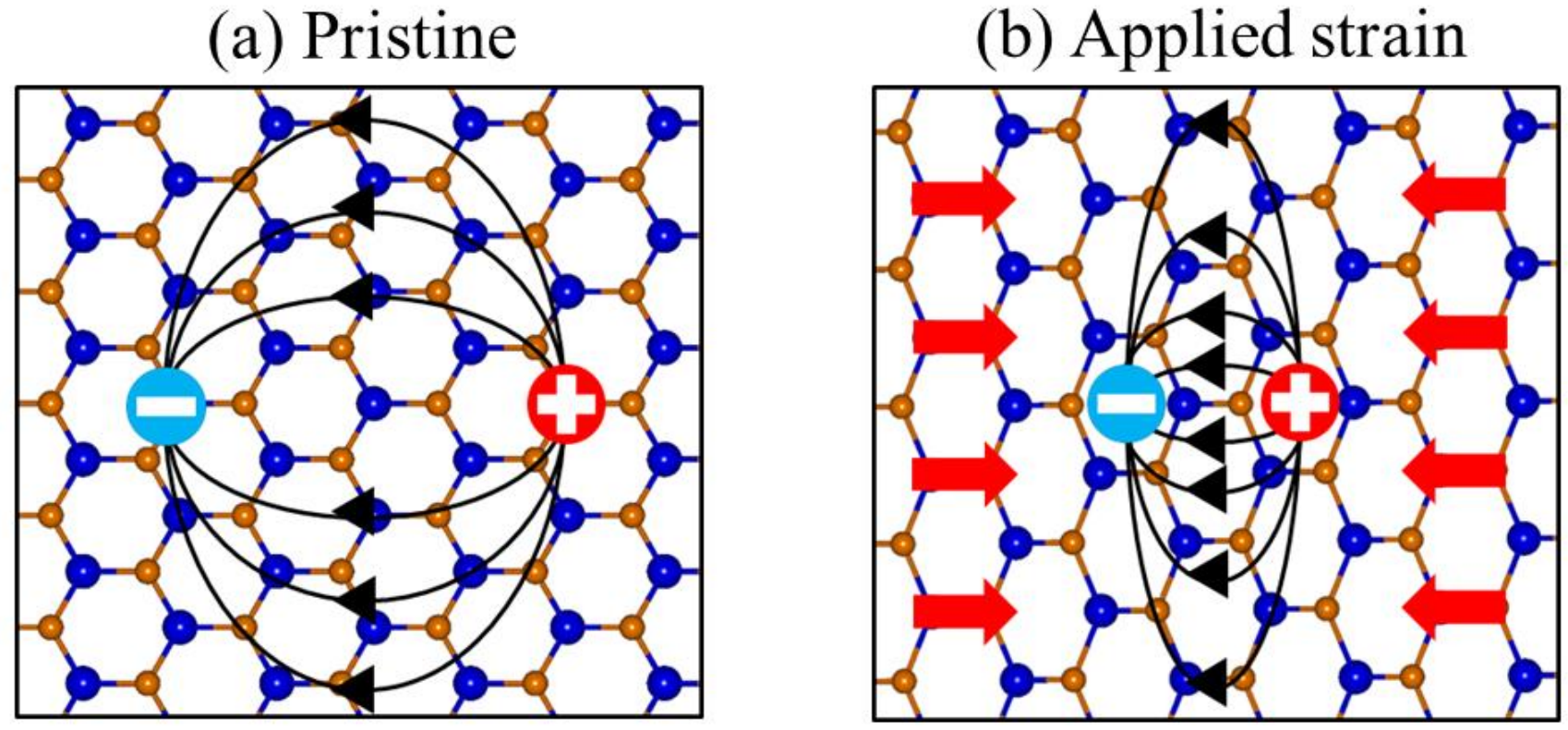

Figure 10. The schematic representation for the mutual couplings of the excited hole and excited electron of monolayer GaSe without (a) and with (b) compressive strain are presented.

When the electromagnetic fields incident into the surface of a medium, some of them will be reflected into the vacuum, while part of them will be transmitted, and part of them will be absorbed and transferred into the thermal reservoirs (Figure 11a). To further comprehend the effects of external mechanical perturbation on the optical properties of the GaSe monolayer, we calculated the absorbance $A(\omega)$, reflectance $R(\omega)$, and transmission $T(\omega)$ spectra. For the normal incidence $\theta = 0^o$ the optical properties $A(\omega)$, $R(\omega)$, and $T(\omega)$ could be expressed as the following equations [73]:

$$A(\omega) = \frac{Re\tilde{\sigma}}{|1 + \tilde{\sigma}/2|^2},$$

$$R(\omega) = \left|\frac{\tilde{\sigma}/2}{1 + \tilde{\sigma}/2}\right|^2,$$

$$T(\omega) = \frac{1}{|1 + \tilde{\sigma}/2|^2},$$

where, $\tilde{\sigma}(\omega) = \sigma_{2D}(\omega)/\varepsilon_0 c$ is the normalized conductivity.

The GaSe monolayer only absorbs ~2% to 4% of the incoming light in case excluding the electron-hole interactions (See Figure S4), most of the incident light is transmitted, while a small portion of them will be reflected on the front surface. However, due to the significant excitonic effects, $A(\omega)$ is enhanced with more than 20% of the incoming light absorbed for the lowest bright exciton states (Figure 11b). The absorption ability of the GaSe monolayer is better than graphene (less than 5%) [73] and $MoS_2$ (15%) [74]. The external mechanical perturbation exerts an important influence on the optical spectra, in addition to the changing of the absorption, reflectance, and transmission intensities, the optical spectra also expressed a significant shift into the visible and the ultraviolet spectral region, respectively, for the elongation and compression. Because the optical absorption coefficient and absorption spectral range can be easily turned by homogeneous and inhomogeneous mechanical modifications, strain engineering has emerged as an effective tool for tailoring the electronic and optical properties of two-dimensional materials for specific application demands, such as electronic, optoelectronic, and photovoltaic devices.

Because of the strong light-matter coupling, the transition energy $E_X^{n=1}$ of the exciton could be achieved by optical methods such as photoluminescence, reflectance, and absorbance spectroscopies. The exciton binding energy $(E_{xb})$, on the other hand, can be evaluated by the energy difference between the exciton groundstate energy $(E_X^{n=1})$ and the fundamental band gap $(E_g)$, $E_{xb} = E_X^{n=1} - E_g$, the latter could be directly measured by scanning tunneling spectroscopy (STS) [59]. Such measurements have been established for GaSe monolayer and our theoretical predictions are in good accordance with their measurements [75].

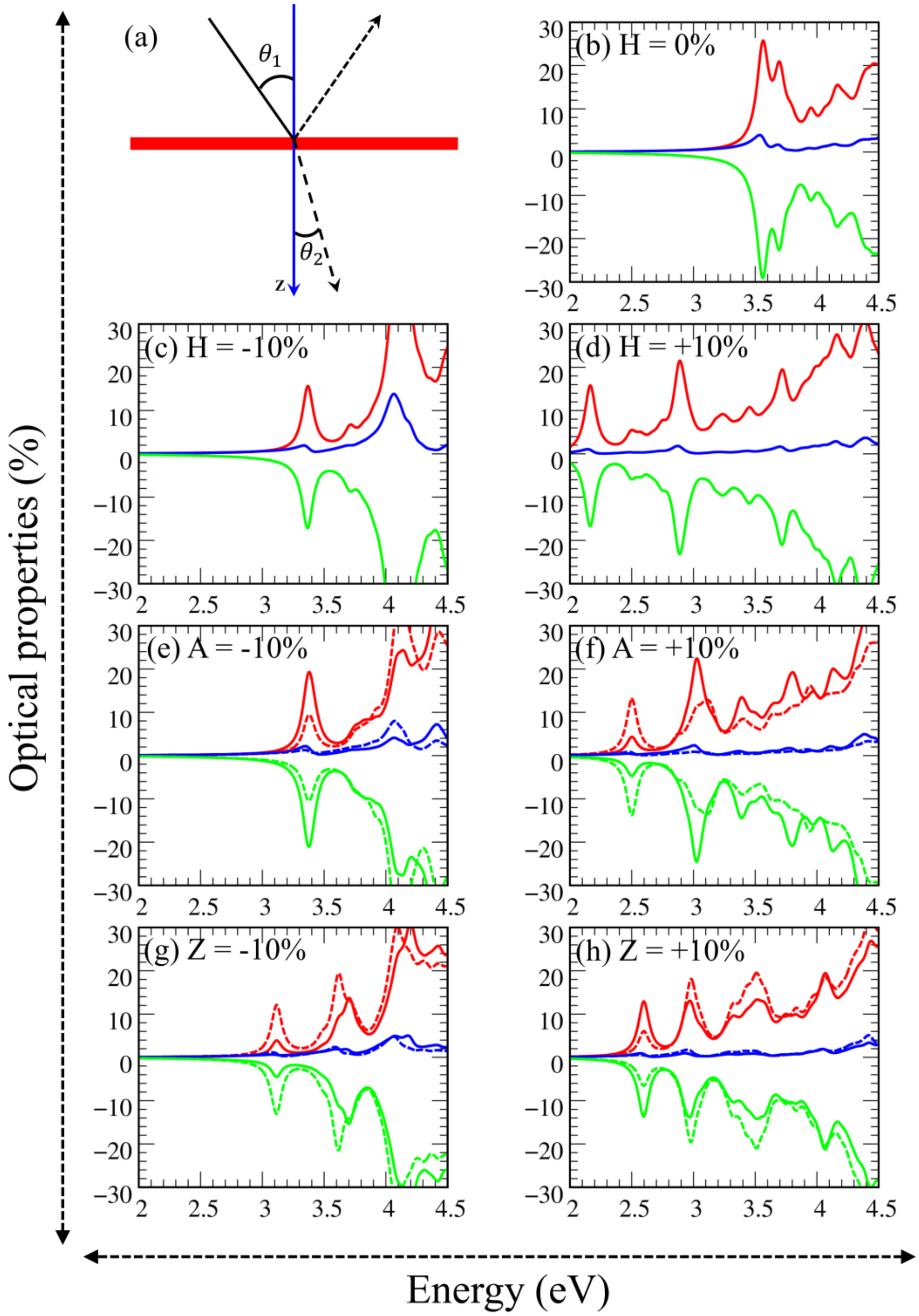

Figure 11. (a) Light propagation in a system consisting of a 2D sheet. The direction of the arrows illustrates incident, reflected, and transmitted light. Frequency-dependent of the optical properties R (blue line), T − 1 (green line), and A (red line) with include excitonic effects for (b) pristine GaSe monolayer, (c) and (d) compressive and tensile homogeneous strains, (e) and (f) compressive and tensile armchair

strains, (g) and (h) compressive and tensile zigzag strains. The solid and the dash lines indicate the polarization of the fields along the armchair and zigzag directions, respectively.

# 4 Conclusions

We investigate the electronic and optical properties of deformed single-layer GaSe using first-principles calculations. The orbital interactions of Ga-4s and Se-$4p_z$ are responsible for the tailoring of $V_1$ and $C_1$ electronic states at Γ, while the in-plane orbital interactions of Ga-($4p_x$, $4p_y$) and Se-($4p_x$, $4p_y$) related to the changing of $C_2$ state at K and $V_2 + V_3$ states at Γ. The electronic band gap and the threshold frequency exhibit considerable red shifts upon the elongation, while these variations are more complicated for compression. Because of the reduced dimensionality and weak dielectric screening, the GaSe monolayer exhibits significant excitonic effects. Our predictions also indicated that the physical distance between two opposite-charge particles controlled by mechanical strains also plays an important role in determining the exciton binding energy, $E_{xb}$ is approximately 0.48 eV for the pristine GaSe, and it decreases (increases) to 0.39 eV (0.493 eV) for the applied homogeneous tensile strain (compressive strain). The absorption coefficient is enhanced due to the presence of strong excitonic effects, the absorbance, reflection, and transmission spectra are also sensitive to external mechanical changes. Similar phenomena are observed for GaSe monolayers under inhomogeneous zigzag and armchair strains; however, due to differences in levels and types of mechanical modification, the electronic and optical spectra present differ slightly in specific characteristics, such as anisotropic energy band structure, non-uniform electronic screening environment, strong polarization of optical absorption spectra, and strain-dependent exciton binding energy. The homogeneous and inhomogeneous mechanical strains appear as a convenient way to turn the electronic and optical properties of GaSe monolayer for specific application demands, e.g., electronic, optoelectronic, and photovoltaic devices.

# Competing interests

The authors declare no competing interests.

# Acknowledgments

This work was financially supported by the Hierarchical Green-Energy Materials (Hi-GEM) Research Center from The Featured Areas Research Center Program within the framework of the Higher Education Sprout Project by the Ministry of Education (MOE) and the Ministry of Science and Technology (MOST 110-2634-F-006 -017) in Taiwan.